# Spectral compression by phase doubling in second harmonic generation

XIN ZENG,[1,2,] SHUZHEN CUI,[1] XIN CHENG,[1] AND YAN FENG[1,2,*]

[1]*Shanghai Institute of Optics and Fine Mechanics, Chinese Academy of Sciences, and Shanghai Key Laboratory of Solid State Laser and Application, Shanghai 201800, China*
[2]*Hangzhou Institute for Advanced Study, University of Chinese Academy of Sciences, Hangzhou, 310024, China*
*\*Corresponding author: feng@siom.ac.cn*

**Abstract:** In second harmonic generation, the phase of the optical field is doubled which has important implication. Here the phase doubling effect is utilized to solve a long-standing challenge in power scaling of single frequency laser. When a ($-\pi/2$, $\pi/2$) binary phase modulation is applied to a single frequency seed laser to broaden the spectrum and suppress the stimulated Brillouin scattering in high power fiber amplifier, the second harmonic of the phase-modulated laser will return to single frequency, because the ($-\pi/2$, $\pi/2$) modulation is doubled to ($-\pi$, $\pi$) for the second harmonic. A compression rate as high as 95% is demonstrated in the experiment limited by the electronic bandwidth of the setup, which can be improved with optimized devices.

High power single frequency lasers at various visible or ultraviolet wavelengths have vital importance for various applications in scientific research and technology development [1–5]. Frequency mixing of near-infrared fundamental lasers is an effective approach. The power scaling of single frequency lasers at visible and ultraviolet wavelength is commonly limited by the corresponding fundamental lasers. Fiber amplifiers, including rare earth doped and Raman fiber amplifiers, are now standard devices to boost continuous-wave near-infrared lasers [6,7]. But in the case of single frequency or narrow linewidth laser, fiber amplifiers are vulnerable to stimulated Brillouin scattering (SBS) effect, which limits their output [8–10].

Phase modulation is an elegant technique and widely used for SBS suppression by broadening the spectrum in a highly controllable way [11–13]. Nevertheless, the amplifier output is not single-frequency anymore. To meet the requirement in applications, method of transforming the phase modulated laser back to single frequency is pursued. The idea behind the phase modulation-demodulation approach is much like the chirped pulse amplification for ultrafast pulsed lasers proposed by Strickland and Mourou [14], which has been successfully and widely applied to solve the problem of nonlinear optical damage in amplifiers [15]. One method of phase demodulation is to do it actively by another phase modulator after the amplifier [10,16]. But the power resistance of commercially available electro-optic modulator is limited to tens of watts.

One important consequence of nonlinear optical interaction is phase modulation [17–19]. For ultrashort pulses, self phase modulation due to the 3rd order nonlinearity is widely used for spectral broadening or compression depending on the initial pulse chirp [19]. For noisy multimode laser, Turitsyn *et al.* reported self-parametric amplification effect which resulted in spectral narrowing in normal dispersion fiber [20]. By synchronous amplitude and phase modulation on an input laser, Goodno *et al.* demonstrated spectral compression after fiber amplifier due to nonlinear self phase modulation [21]. But the resulting laser is amplitude modulated, which is undesirable for many precision applications.

In this paper, we introduce a passive spectral compressing method utilizing the phase doubling effect in second harmonic generation (SHG). When a discrete phase modulation of $\pi$ difference is applied to a single frequency laser, the frequency doubled laser will return to single frequency output because the phase modulation is doubled in SHG. Typical binary phase modulation of square wave and pseudo-random binary sequence (PRBS) are investigated both theoretically and experimentally. A compression rate as high as 95% is demonstrated in the experiment, which can be improved with optimized devices of higher bandwidth.

The concept is illustrated in Fig. 1. A single frequency seed laser is phase modulated to have a broader spectrum before sending into a fiber amplifier, so as to reduce the spectral density and suppress SBS in the amplifier. After SHG, single-frequency frequency doubled laser can be generated from the amplified broadband laser if proper phase modulation is applied. Considering a phase modulated laser, the optical field strength is represented as [19]

$$\tilde{E}(t) = E e^{-i(\omega t + \varphi(t))} + \text{c.c.} \quad (1)$$

where $E$ is field amplitude, $\omega$ is the laser frequency, and $\varphi(t)$ is the phase modulation function. The second order nonlinear polarization is given by

$$\tilde{P}^{(2)}(t) = \epsilon_0 \chi^{(2)} \tilde{E}^2(t) = 2\epsilon_0 \chi^{(2)} E E^* + \left( \epsilon_0 \chi^{(2)} E^2 e^{-i(2\omega t + 2\varphi(t))} + \text{c.c.} \right) \quad (2)$$

where the first term leads to optical rectification, and the second term leads to the generation of radiation at the second harmonic frequency. Generally, as seen in Eq. (2), the second harmonic has a phase modulation term which is twice of that on the fundamental laser. Here we consider a special case where the phase of the fundamental laser is modulated alternatively between two values of $\pi$ difference.

$$\varphi(t) = -\pi/2 \text{ or } \pi/2 \quad (3)$$

Then the second harmonic will have a phase modulation of $2\varphi(t) = -\pi$ or $\pi$, which is of $2\pi$ difference and effectively no modulation. In other words, a phase modulated single frequency laser is demodulated to single frequency in second harmonic generation.

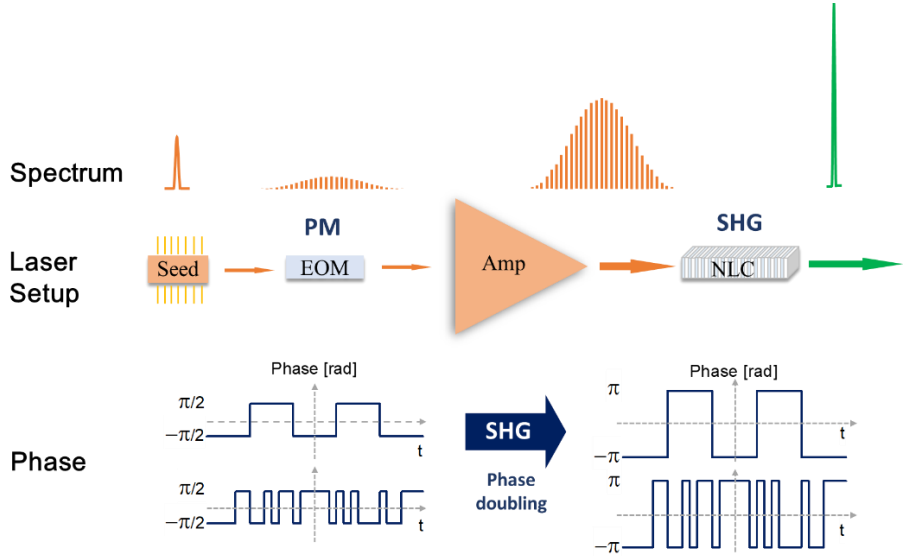

Fig. 1. Illustration of the concept. A single frequency seed laser is phase modulated to have a broader spectrum before sending into a fiber amplifier. After SHG, single frequency laser can be generated from the amplified broadband laser if discrete phase modulation of $\pi$ difference is applied, because the phase is doubled to $2\pi$ modulation after SHG. The cases of square wave and PRBS modulation are shown.

Experimental investigation were carried out on a setup which consists of a phase modulated single frequency seed laser, two stages of fiber amplifier, a SHG unit, and laser characterization devices, as shown in Fig.2. The single frequency seed laser is a fiber-pigtailed distributed feedback (DFB) diode laser at 1064 nm with a linewidth of 1 MHz and an output power of 30 mW. The DFB laser is protected by an optical isolator, and phase modulated by an electro-optic phase modulator (EOM) of 10 GHz bandwidth. The square wave and PRBS signal are generated from a bit error rate tester (Anritsu MP2100B), which are applied to the EOM after amplification. The two-stage optical fiber amplifier boosts the 1064 nm laser to 13.3 W in linear polarization. After a free space optical isolator, maximum 11.7 W is available for the SHG experiment. The amplified 1064 nm laser is focused with a 60 mm lens into a 25-mm periodically poled lithium niobate crystal (PPLN), where the second harmonic light at 532 nm is generated. The PPLN crystal is kept in an oven at a temperature of 67.62 $^0$C with a stability of ± 0.01 $^0$C for phase matching. The spectral acceptance bandwidth of the PPLN in the experiment is calculated to be about 0.083 nm (22 GHz) [23], which is significantly wider than the laser linewidth. A second harmonic output of about 1 W was obtained with a conversion efficiency of 9% at different modulation conditions. The optical spectra of the 1064 nm fundamental laser and the 532 nm second harmonic were measured with two Fabry-Perot Interferometers (FPI) at corresponding spectral range.

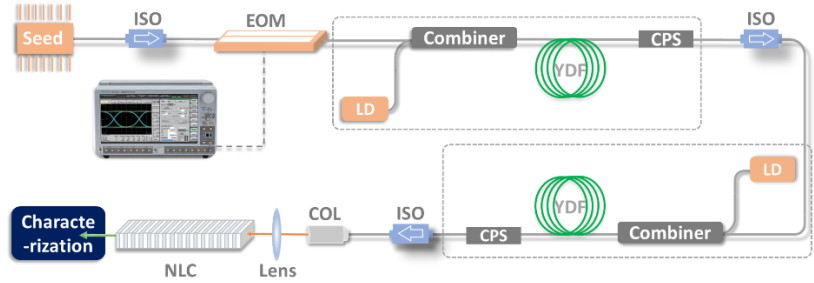

Fig. 2. The experimental setup consists of a single frequency diode seed laser, a phase modulator, two stages of fiber amplifier, a SHG unit, and laser characterization devices. ISO: isolator; EOM: electro-optic modulator; LD: laser diode; YDF: Ytterbium doped fiber; CPS: cladding power stripper; COL: collimator; NLC: nonlinear crystal.

Square wave and PRBS are important examples of binary phase modulation [11]. The square wave modulation is one of the simplest form, which is intuitive for understanding the process.

With square wave modulation, discrete sidebands are generated at both side of the carrier wave ω. The separation between the frequency components are determined by the modulation frequency Ω. The power distribution among the carrier wave and sidebands are determined by the modulation depth γ. Fig. 3(a) shows the simulated spectra of the fundamental laser and corresponding second harmonic for γ = 0, π/2, and π. γ = 0 corresponds to the unmodulated single frequency case. When γ = π/2, the modulation depth on the second harmonic is π. The power spectral density at carrier frequency becomes null, which is characteristic for π modulation. When γ = π, the modulation depth on the second harmonic is 2π, which equals to 0 modulation. Therefore, the phase modulation generated few frequency laser is converted back to single frequency as a second harmonic output.

Fig. 3(b) plots the corresponding experimental results with a modulation frequency Ω = 50 MHz. The spectra were measured with two FPIs of 4 GHz free spectral range for fundamental laser and frequency doubled laser, respectively. Due to the limited finesse of FPIs, the spectral lines have finite width. The spectral evolution of the second harmonic generation process behaves as predicted by the simulation. With π step square wave modulation, the second harmonic is compressed to single frequency. The spectra are normalized to the single frequency case. It is found that the height of the demodulated single frequency second harmonic is 95.2% of the unmodulated single frequency output. We define this ratio as compression rate. The factors which influence the compression rate are discussed later.

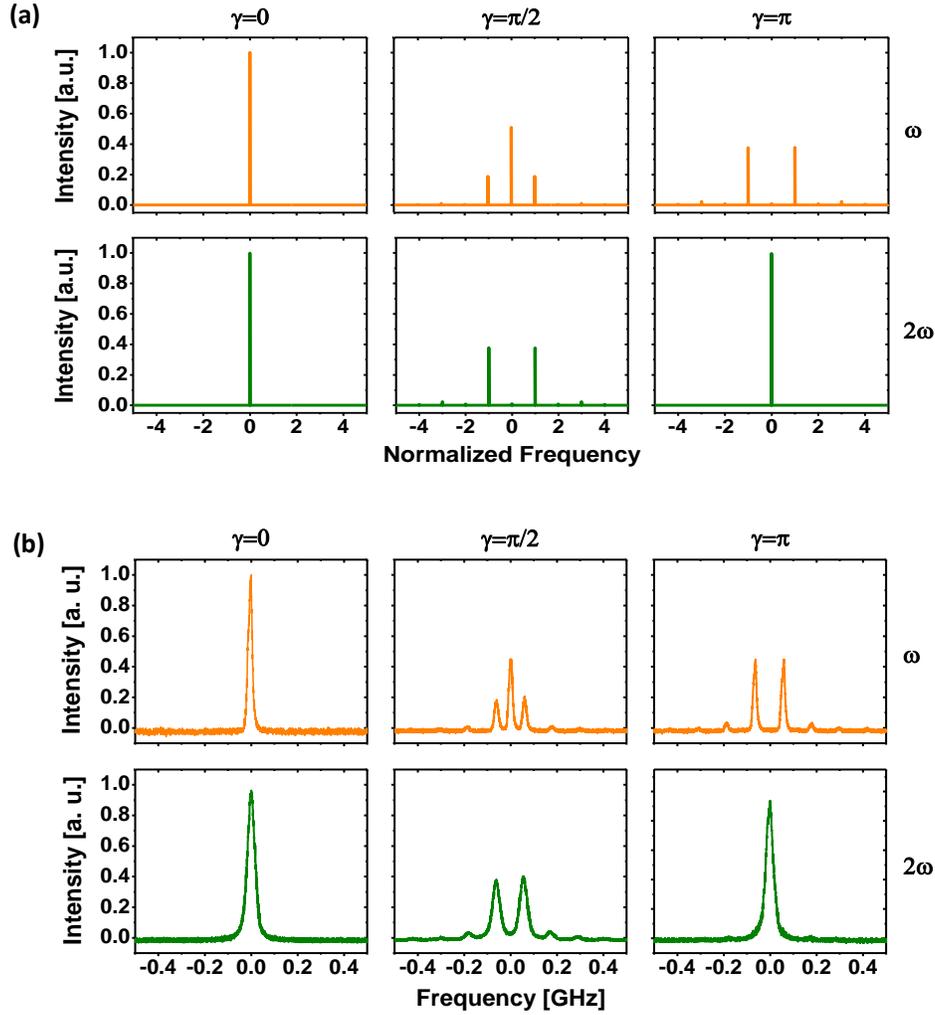

Fig. 3. Spectral evolution in the case of square wave phase modulation. (a) Simulated spectra of the fundamental laser (up panel) and corresponding second harmonic (bottom panel) for modulation depth γ = 0, π/2, and π, respectively, in the case of square wave modulation. The frequency displacement is normalized to the modulation frequency. (b) corresponding experimental results with a modulation frequency Ω = 50 MHz.

In practice, PRBS modulation is more commonly used in the development of narrow linewidth fiber amplifier [11–13]. One reason is that lower spectral density can be generated, which is critical for SBS suppression and power scaling of the amplifier output. The spectral compression method also works with PRBS modulation, as long as the modulation depth is $\pi$. Fig. 4(a) shows the simulation results for the case of PRBS7 (PRBS with shift register length of 7). As compared to square wave modulation, much more frequency components are generated. Therefore, the peak spectral density of the fundamental laser is reduced significantly by a factor of approximately $2^7$=128. After second harmonic generation, the laser is again converted back to single frequency.

A variety of experiments were carried out with different bit rates and shift register lengths. Results with PRBS7 of 200 Mbps bit rate are shown in Fig. 4(b). Again, the spectral evolution of the second harmonic generation process follows the theoretical prediction well. It should be noted that the higher height of the broadband spectral components as compared to the numerical simulation is due to the spectral resolution in the experiments. The spectral resolution of the

FPIs for the fundamental laser and second harmonic are about 21 MHz and 43 MHz, respectively. But the spectral separation between adjacent components is $200 / (2^7-1) = 1.57$ MHz. Therefore, the relative height of the broadband spectra is increased artificially by a factor of approximately 13 and 27 for fundamental laser and second harmonic, respectively. A compression rate of 82.4% is measured. The uncompressed part of second harmonic forms the pedestal of the laser spectrum.

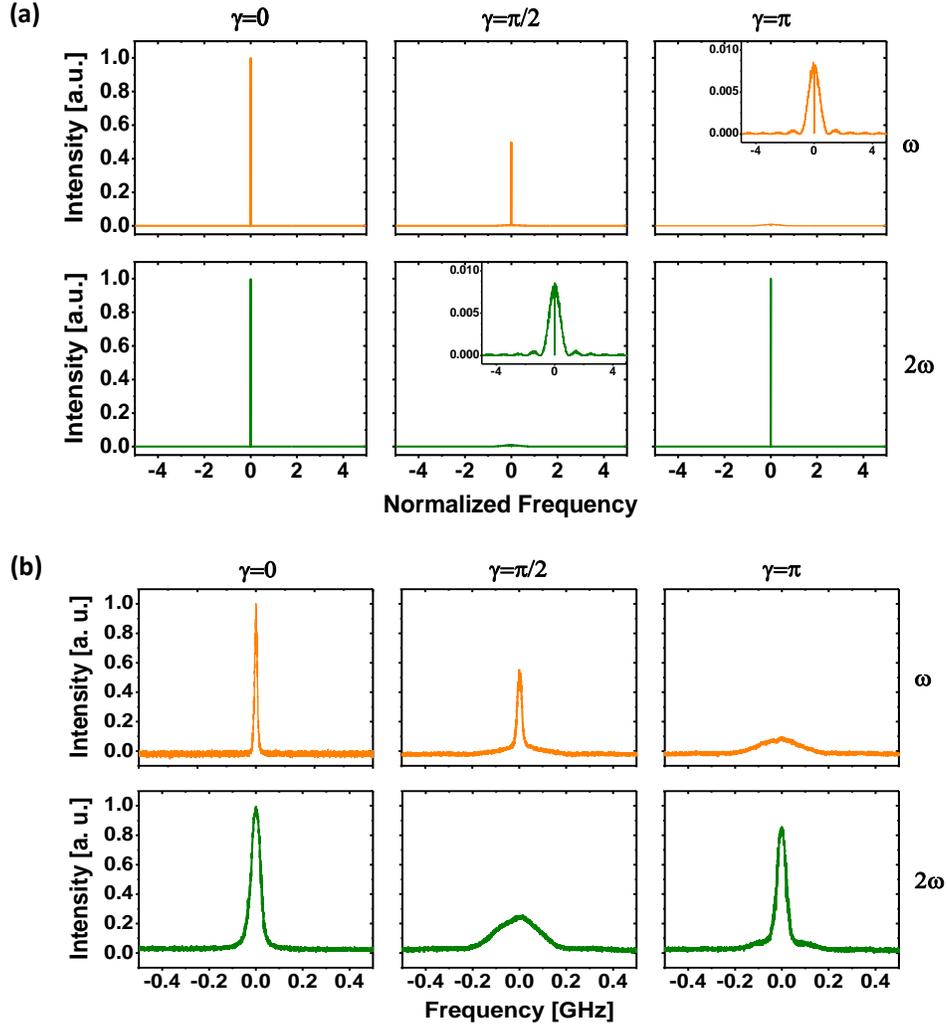

Fig. 4. Spectral evolution in the case of PRBS7 phase modulation. (a) simulated spectra of the fundamental laser (up panel) and corresponding second harmonic (bottom panel) for modulation depth $\gamma = 0$, $\pi/2$, and $\pi$, respectively, in the case of PRBS7 modulation. The frequency displacement is normalized to the bit rate. Insets are the zoom-in view of the corresponding spectra. (b) corresponding experimental results with a PRBS7 modulation of 200 Mbps bit rate.

Thus, we have shown that the $\pi$ step phase modulated laser can indeed be converted back to single frequency as a second harmonic output. It provides an effective method to obtain high power visible or ultraviolet single frequency laser after frequency doubling. The method is useful because visible or ultraviolet single frequency laser are commonly generated by frequency doubling of longer wavelength fundamental laser. A measure of the effectiveness of the method is the compression rate. The compression rate with respect to bit rate for PRBS7 modulation were investigated and the results are shown in Fig. 5(a). The demodulation rate decreases gradually with respect to bit rate.

With ideal (0, $\pi$) binary phase modulation, the demodulation rate should be 100% according to the theory. However, in real world experiment, the phase modulation is always deviated from the ideal case. First of all, the phase modulator has a limited response time, which is characterized as the bandwidth. Secondly, the modulation signal applied on the modulator is also not an ideal two-value function, because the pattern generator, the modulator driver, and the electronic cables all have their finite bandwidth. All these factors contribute to the lower-than-one demodulation rate in the experiment. In the experiments, we have just used the devices and components available in our lab. The compression rate can be improved with devices of higher bandwidth, which are available in the market.

Figure 5(b) shows the output power of second harmonic generation with respect to the PRBS7 bit rate. With higher bit rate, the linewidth is broadened wider, as much as approximately 800 MHz at 800 Mbps, which means higher SBS suppression capability. Because the broadened linewidth is still far within the spectral acceptance of the nonlinear crystal used in the experiment (22 GHz), SHG power keeps constant with varying bit rate, and is the same as that in the unmodulated single frequency case.

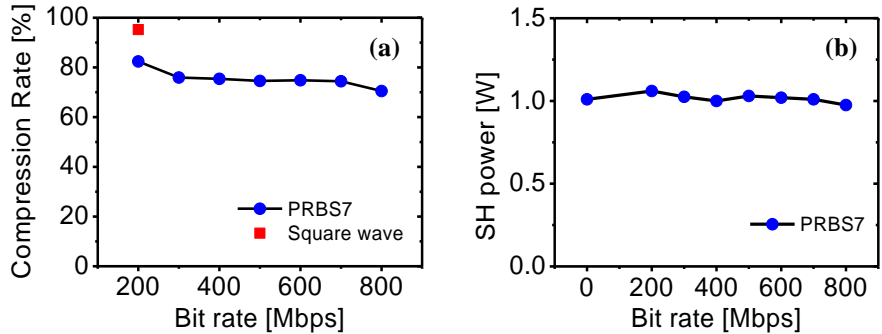

Fig.5: Compression rate and second harmonic output. (a) Compression rate with respect to bit rate for PRBS7 modulation and compression rate for square wave modulation with a bit rate of 200 MHz and a frequency of 50 MHz. (b) output power of second harmonic with respect of the PRBS7 bit rate.

In conclusion, we have introduced and demonstrated the spectral compression by phase doubling in second harmonic generation process. Compression rate as high as 95% is demonstrated in the experiment for square wave modulation. Around 80% compression rate is demonstrated with PRBS modulation with the devices and components available in our lab. The performance can be improved with optimized setup with higher electronic bandwidth. Since PRBS phase modulation is widely used in high power narrow linewidth fiber amplifiers and up to kilowatt SHG has been demonstrated with such laser sources [11,12,22], the proposed method may enable a leap in power scaling of single frequency visible and ultraviolet lasers.

Furthermore, it should also be noted that it is straightforward to apply similar phase manipulation in other wave mixing processes, including sum frequency generation (phase addition), difference frequency generation (phase subtraction), four wave mixing (phase addition and subtraction), etc. Therefore, a variety of phase and spectral manipulation can be done in wave mixing process, which opens a new field of development in nonlinear optics and laser technology.

**Funding.**

This work was supported by The National Key R&D Program of China (2020YFB04012600, 2020YFB1805900, 2018YFB0504600); National Natural Science Foundation of China (No. 62075226); Shanghai Science and Technology Commission (19441909800).